\documentclass{ws-procs9x6}
\usepackage{graphicx,rotating_pr}
\begin{document}
\title{Past, Present and Future of Muonium}
\author{Klaus P. Jungmann }
\address{Kernfysisch Versneller Instituut\\
Rijksuniversiteit Groningen\\
Zernikelaan 25, \\ 
Groningen, 9747 AA, The Netherlands\\ 
E-mail: jungmann@KVI.nl} 
\maketitle
\abstracts{Muonium, the atom which consists of a positive muon
and an electron, has been discovered by a team led by Vernon W. Hughes
in 1960. It is in many respects the most ideal atom available from nature.
Due to the close confinement in the bound state muonium can be used as an 
ideal probe 
of electro-weak interaction, including particularly  
Quantum Electrodynamics, and to 
search for
additional yet unknown interactions acting on leptons. Recently completed 
experiments
cover the ground state hyperfine structure, the 1s-2s interval and a search for
spontaneous conversion of muonium to antimuonium. The experiments yield
precise values for the fine structure constant, the muon mass  
and its magnetic moment. The results from these precision measurements 
have provided restrictions for a number of theories 
beyond the Standard Model of particle physics. Future precision experiments
will require new and intense sources of muons.
}

\section{Muonium - The Atom discovered by Vernon Hughes}

Atomic hydrogen is generally considered the simplest and most fundamental
atom in nature. Its role in development of modern physics
is outstanding. Our physical picture of atoms, the success of quantum
mechanics and the start of quantum field theories such as Quantum 
Electrodynamics (QED)
are just a few examples for insights which go back to careful analysis
of what had been observed in this atom.
Hydrogen has been exploited in numerous precision measurements
to determine fundamental constants and to reconfirm fundamental
concepts such as, e.g. the equality of the electron and proton charge units.

\begin{figure}[ht] 
\centerline{\epsfxsize=3.4in\epsfbox{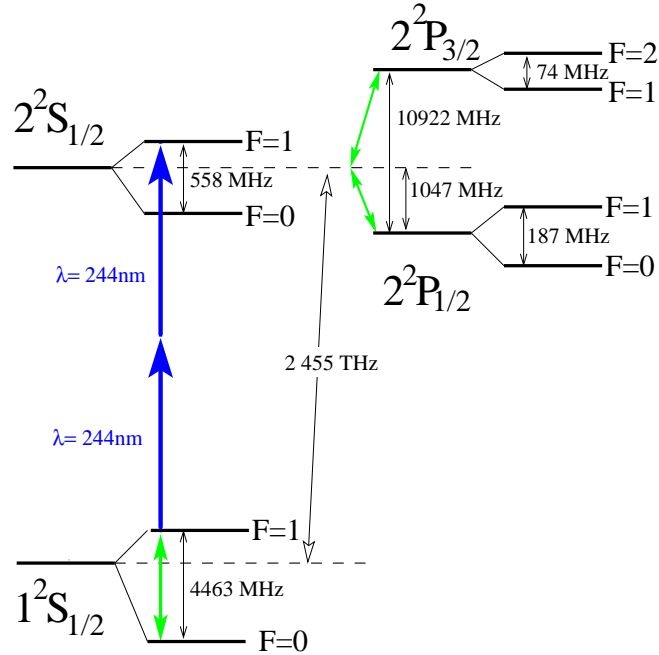}}   
\caption{ \label{M_term} 
 Energy levels of the hydrogen-like muonium atom 
 for states with principal quantum numbers 
 n=1 and n=2.
 The indicated transitions could be induced to date by microwave or 
 laser spectroscopy. 
 High accuracy has
 been achieved for the transitions which involve 
 the ground state. The atoms   can be produced most efficiently for n=1.}
\end{figure}

Unfortunately, the presence of the proton as the nucleus
in this one electron atom reduces the possibilities for a complete
theoretical description. Precise measurements are at present 
orders of magnitude more  accurate than calculations can be performed.  
Proton properties such as its  mean square charge radius or its 
magnetic radius 
or even the dynamics  of the charge and spin carrying constituents 
inside the proton
are not known to sufficient accuracy.

High energy scattering experiments have shown for leptons no structure down to
dimensions of $10^{-18}$~m. They may therefore be considered ''point-like''. 
As a consequence, complications as those arising from the structure of the 
nucleus
in natural atoms and such artificial systems that contain hadrons are
absent in the muonium atom (${\rm M}=\mu^+ e^-$), which is the bound 
state of two leptons, 
a positive muon ($\mu^+$)  and an electron ($e^¯$)
\cite{Hughes_1990,Jungmann_1999}. It may be considered a 
light hydrogen isotope.

The dominant interaction within the muonium atom  (see Fig. \ref{M_term}) 
is electromagnetic. In the framework of bound state Quantum 
Electrodynamics (QED) the electromagnetic part of the binding
can be calculated to sufficiently high 
accuracy for  modern high precision spectroscopy experiments. There are also 
contributions from weak interactions arising through $Z^0$-boson exchange 
and from
strong interactions owing to vacuum polarization loops containing 
hadrons. The corresponding energy level shifts
can be obtained to the required level of precision using 
standard theory. Precision experiments in muonium can therefore provide
sensitive tests of the Standard Model of particle physics 
and sensitive searches for new and yet unknown 
forces in nature become possible.
Parameters in speculative theories can be restricted. In particular, such 
speculations
which try to expand the Standard Model in order to  gain deeper insights 
into some of its not well understood features, i.e. where the 
standard theory gives well a full description, but lacks a
fully satisfactory explanation of the observed facts. 
In addition, fundamental constants like the muon mass $m_{\mu}$, its magnetic 
moment 
$\mu_{\mu}$ and magnetic anomaly $a_{\mu}$ and the 
fine structure constant $\alpha$
 can be measured precisely by muonium spectroscopy.   


In 1960 a team led by Vernon W. Hughes has observed the muonium atom for the
first time \cite{Hughes_1960}. 
The details of the exciting circumstances around this discovery and 
the research in the early  years are described in this volume by an eye 
witness, Richard Prepost \cite{Prepost_2004}. 
They are also available from the viewpoint of the leader,
Valentin Telegdi,
of the very group, which was competing in muonium research 
with the Yale team of Vernon Hughes for more than a decade \cite{Telegdi_1991}.

\section{Muonium Formation} 

\begin{figure}[ht] 
\centerline{ \rotatebox{-90}{\epsfxsize=\textwidth\epsfbox{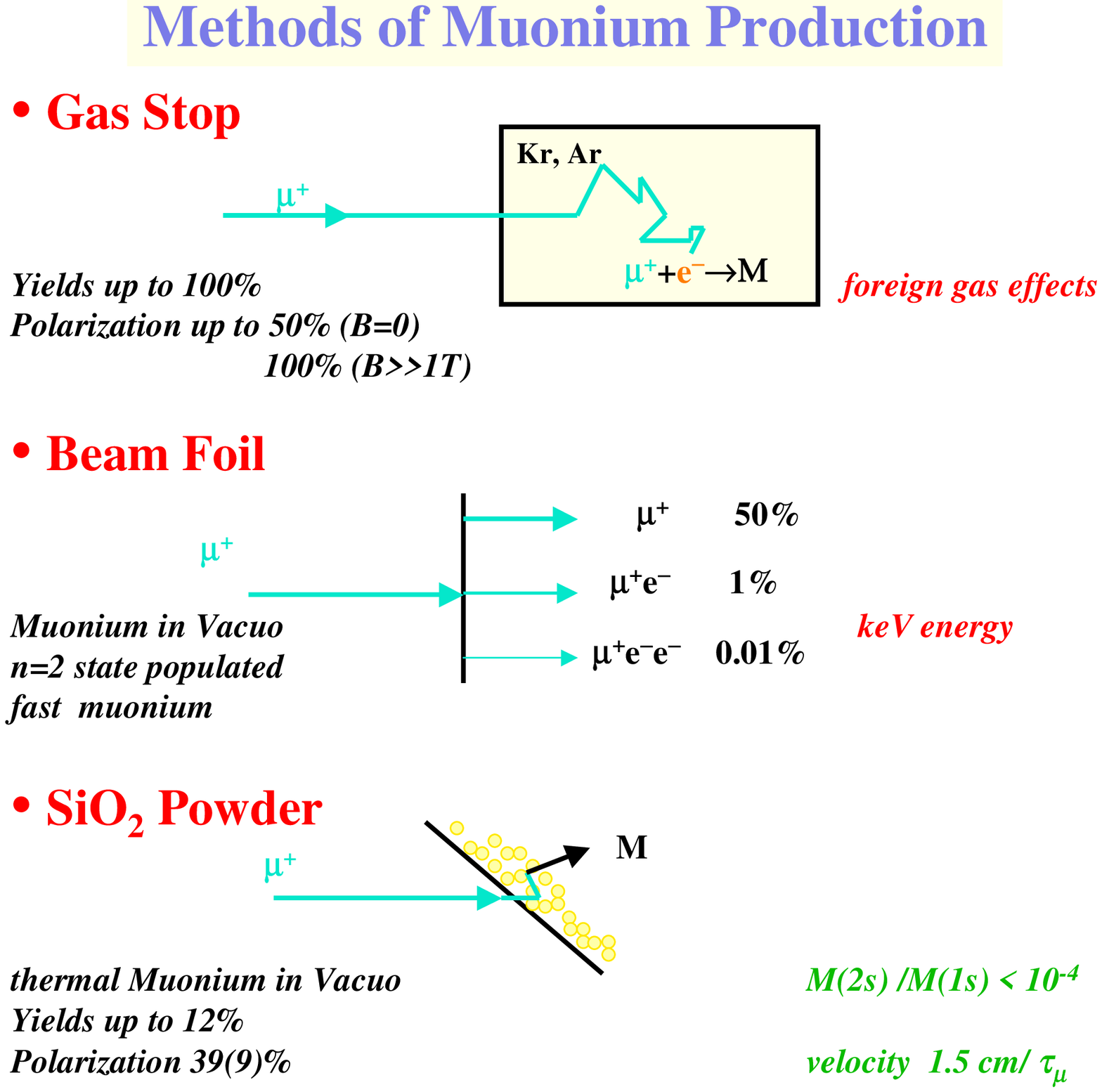}}}   
\caption{\label{mprod}
Muonium atoms for precision experiments have been produced by
three different methods. Stopping muons in a noble gas gives atoms
at thermal energies which are subject to collisional effects. Due to 
velocity a beam foil technique yields muonium atoms in vacuum
at keV energies. Muonium atoms diffuse at thermal velocities in
vacuum after being produced by muon stopping in a fluffy SiO$_2$ powder.}
\end{figure}

In  the early years muonium research concentrated on
measurements that were possible with
atoms created by stopping muons in a material
and studying them in this environment (see Fig. \ref{mprod}).
Besides important work on condensed
matter in the framework of muon spin rotation ($\mu$SR)
there were in particular
precision experiments which concerned the ground state hyperfine 
structure and a search for muonium-antimuonium conversion. 
In the 1980ies the spectrum of possible experiments
could be significantly expanded when methods
where developed which allowed to have the 
atoms in vacuum \cite{Hughes_1990,Jungmann_1999}.

All high precision experiments in muonium up to date atom 
have involved the 1s ground state (see Fig.\ref{M_term}),
in which the atoms can be produced in sufficient quantities. 
The most efficient 
mechanism is $e^-$ capture after stopping $\mu^+$ in a suitable noble gas,
where yields of 80(10)~\% were achieved for krypton gas. This technique 
was used in
the most recent precision measurements of the atom's ground state hyperfine 
structure splitting
$\Delta \nu_{HFS}$ and $\mu_{\mu}$
at the Los Alamos Meson Physics Facility (LAMPF) in Los Alamos, USA
\cite{Liu_1999}. 

Muonium at thermal velocities in vacuum can be obtained
by stopping $\mu^+$ close to the surface of a SiO$_2$ powder target, where 
the atoms are formed through $e^-$ capture and some of them diffuse through 
the target surface
into the surrounding vacuum. This process has an efficiency of a few percent 
and
was an essential prerequisite for Doppler-free two-photon laser spectroscopy
of the 1$^2$S$_{1/2}$-2$^2$S$_{1/2}$ interval $\Delta \nu_{1s2s}$ at the 
Rutherford Appleton Laboratory (RAL) in Chilton, United Kingdom 
\cite{Meyer_2000},
which yields an accurate value for $m_{\mu}$.

Electromagnetic transitions in excited states, particularly 
the 2$^2$S$_{1/2}$-2$^2$P$_{1/2}$
classical Lamb shift and 2$^2$S$_{1/2}$-2$^2$P$_{3/2}$ fine structure
splitting could be induced by microwave spectroscopy, too. 
Only moderate
numbers of atoms in the metastable 2s state can be produced
with a beam foil technique. 
Because furthermore these atoms have keV energies due
to a velocity resonance in their formation, 
the experimental accuracy is now the 1.5~\% level \cite{Ora_84,Bad_84},
which represents not yet a severe test of theory. 

\section{Muonium Ground State Hyperfine Structure}

\subsection{The Last LAMPF Experiment}

\begin{figure}[ht] 
\centerline{\epsfxsize=4.1in\epsfbox{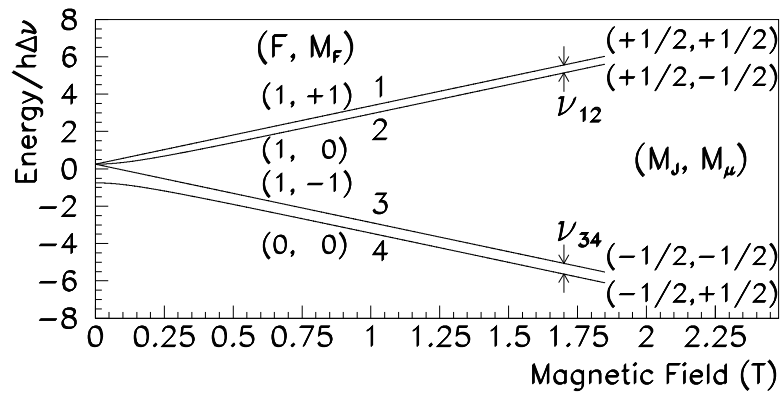}}   
\caption{ \label{BrRa}
The muonium ground state Zeeman splitting.  }
\end{figure}

\begin{figure}[ht] 
\centerline{\epsfxsize=4.2in\epsfbox{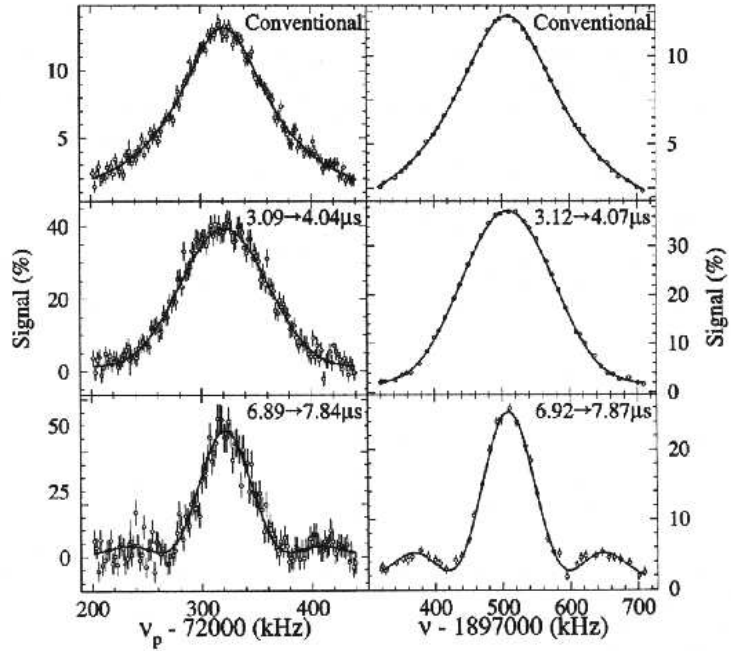}}   
\caption{  \label{HFS_sig}
Samples of conventional and ''Old Muonium'' resonances at frequency 
$\nu_{12 }$.  
The narrow ''old'' lines exhibit a larger signal amplitude.  
The signals were obtained with magnetic field sweep (left column, 
magnetic field 
in units of proton NMR frequencies) and by
 microwave frequency scans (right column).}
\end{figure}

The most recent experiment at LAMPF had a krypton gas target inside of a microwave 
cavity at typically atmospheric density and in a homogeneous magnetic field of 
1.7 T.
Microwave transitions between the two energetically highest 
respectively two lowest Zeeman 
sublevels of the n=1 state at the frequencies $\nu_{12}$ and $\nu_{34}$
 (Fig. \ref{BrRa}) 
involve 
a muon 
spin flip. 
Due to parity violation in the weak interaction muon decay process
 the 
$e^+$ from $\mu^+$ decays are preferentially emitted in the $\mu^+$ spin 
direction. This allows a detection of the spin flips through a change in the 
spatial distribution of the decay $e^+$. As a consequence of the Breit-Rabi 
equation, 
which describes the behaviour of the muonium ground-state Zeeman levels in a
 magnetic 
field  B, the sum of $\nu_{12}$ and $\nu_{34}$
equals at any value of B the zero field splitting $\Delta \nu_{HFS}$. 
For sufficiently well known B the difference of these two frequencies yields 
the magnetic moment $\mu_{\mu}$.
The latest LAMPF experiment \cite{Liu_1999}  has utilized the technique of
 ''Old Muonium'', 
which allowed to reduce the line width of the signals below half of the 
"natural" line width $\Delta \nu_{nat}= 1/(2 \pi \tau_{\mu})$ 
(Fig. \ref{HFS_sig})\cite{Bos_95}, where $\tau_{\mu}$ = 2.2$\mu$s
is the muon lifetime. 
For this purpose an
essentially continuous
muon beam was chopped by an electrostatic kicking 
device into 4 $\mu$s long pulses with 14 $\mu$s separation. 
Only decays of atoms which had been interacting coherently with the microwave 
field for periods longer than several muon lifetimes
were detected.
 Here,  the first quoted uncertainty is due to the accuracy to 

The magnetic moment was measured to be  $\mu_{\mu}$= 3.183\,345\,24(37)\,(120 ppb) which 
translates into a muon-electron mass ratio $\mu_{\mu}/m_e$ = 206.768\,277(24) (120 ppb).
The zero-field hyperfine splitting is determined to 
$\Delta \nu_{HFS}(exp)$ = 4\,463\,302\,765(53) Hz (12 ppb) which
 agrees well with the theoretical prediction of 
$\Delta \nu_{HFS}(theo)$ = 4\,463\,302\,563(520)(34)($<$100) Hz (120 ppb). 

which $m_{\mu}/m_e$ is known, the second error is from the knowledge of a as
 extracted from Penning 
trap measurements of the electron magnetic anomaly, and the third 
uncertainty corresponds to estimates of uncalculated higher order terms. 
Among the non-QED contributions is the strong interaction through vacuum 
polarization loops with hadrons 
 (250 Hz) and a parity conserving axial vector-axial vector weak
 interaction (-65 Hz).
 
For the muonium hyperfine structure the comparison between theory and 
experiment is possible with almost two orders of magnitude higher precision 
than for natural hydrogen because of the not sufficiently known proton charge 
and magnetism distributions. 
The achieved -- some six orders of magnitude higher -- experimental 
precision in hydrogen maser experiments can unfortunately not 
be exploited for a better understanding of fundamental interactions.
Among the possible exotic interactions, which could contribute to 
$\Delta \nu_{HFS}$, 
is muonium-antimuonium  conversion \cite{Willmann_1999}. 
Here, an upper limit 
of 9 Hz could be set from an independent experiment described in section
\ref{MMbar}.
\begin{figure}[hbt] 
\centerline{\epsfxsize=3.1in\epsfbox{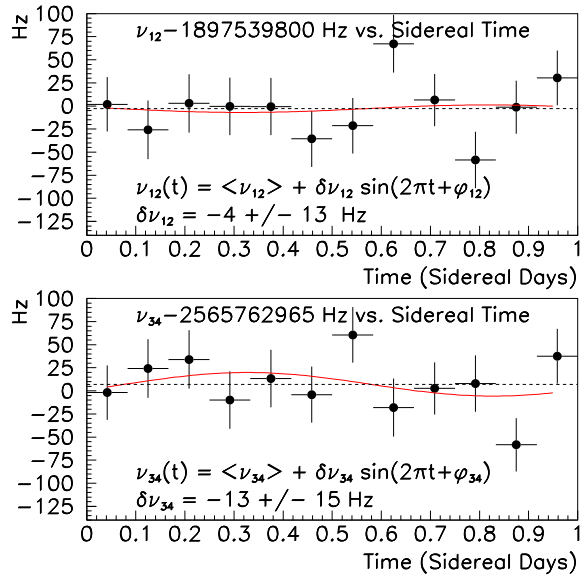}}   
\caption{\label{CPT_fig}
The absence of a significant siderial oscillation
confirms CPT invariance at the best level tested for muons. }
\end{figure} 

\subsection{Search for CPT violation} 

Recently, generic extensions of the Standard Model, 
in which both Lorentz invariance and CPT invariance are not assumed, have 
attracted widespread attention in physics. 

Diurnal variations of the ratio 
($\nu_{12} - \nu_{34})/(\nu_{12} + \nu_{34}$) are predicted (see 
Fig. \ref{CPT_fig}). 
An upper  limit could be set from a 
reanalysis of the LAMPF data at 
$2\cdot 10^{-23}$ GeV for the Lorentz and CPT violating parameter. 
In a specific 
model by Kostelecky and co-workers a dimensionless figure of merit for CPT 
tests is sought by normalizing this parameter to the particle mass. In this 
framework $\Delta \nu_{HFS}$ provides a significantly better test of 
CPT invariance 
than electron g-2 and the neutral
Kaon oscillations\cite{Hughes_2001}.

\subsection{The Fine Structure Constant}

The hyperfine splitting is proportional to $\alpha^2 \cdot R_{\infty}$
with the very precisely 
known Rydberg constant $R_{\infty}$. 
Comparing experiment and theory yields  
$\alpha^{-1}$= 137.035\,996\,3(80) (58ppb). 
If $R_{\infty}$ is decomposed 
into even more 
\begin{figure}[ht] 
\centerline{\epsfxsize=4.2in\epsfbox{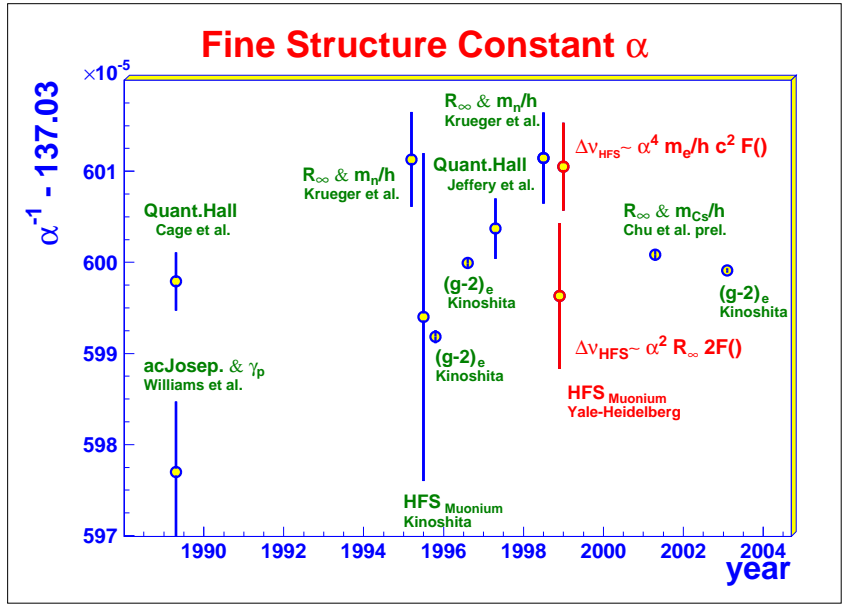}}   
\caption{Various determinations of the fine structure constant $\alpha$.}
\end{figure}
fundamental constants, one finds $\Delta \nu_{HFS} = \alpha^4·m_e/h$.
 With $h/m_e$ as determined in measurements of the neutron de Broglie 
 wavelength  we have $\alpha^{-1}$= 137.036\,004\,7(48) (35 ppb).
 In the near 
 future a small improvement in this figure  can be expected from 
ongoing determinations of $h/m_e$ in measurements of the photon 
recoil in Cs. A better determination of the muon mass, e.g. will 
result in a further improvement and may contribute to resolving the 
situation of various poorly agreeing determinations of the fine structure
constant, which is important in many different fields of physics.

It should be mentioned that the present agreement between $\alpha$ as 
determined from muonium hyperfine structure and from the electron magnetic 
anomaly is generally considered the best test of internal consistency 
of QED, as one case involves bound state QED and the 
other one QED of free particles. 

\subsection{Future Possibilities for $\Delta \nu_{HFS}$}

The results from the LAMPF experiment are mainly statistics limited
and improve the knowledge of both $\Delta \nu_{HFS}$ and $\mu_{\mu}$ 
by a factor of three over previous measurements. 
This gain could be significantly surpassed
with an experiment based on the ''Old Muonium'' technique
at a future high flux muon source (see section \ref{Future}). As a useful 
starting point one would like to
have $5\cdot10^8$ $\mu^+$/s at below 28 MeV/$c$ momentum with typically 
1 \% momentum width. The beam should be pulsed with
1$\mu$s wide pulses of up to several 10 kHz repetition frequency. 
 One can expect that
theory will be continuously improved to allow
the extraction of fundamental physics information from
a precision experiment\cite{Eides_2003}.

\section{Muonium 1s-2s Two-photon Spectroscopy}

In muonium the 1s-2s energy difference is essentially given by the relevant 
quantum numbers, $R_{\infty}$ and a reduced mass correction. Therefore, this transition 
may be regarded ideal for a determination of the muon-electron mass ratio. QED 
corrections are well 
known for the needs of presently possible precision experiments and do not play 
an important role here.

Doppler-free excitation of the 1s-2s transition has been achieved in pioneering 
experiments at KEK \cite{Chu_88}  and at RAL \cite{Maas_94}. 
In all these experiments two counter-propagating 
pulsed laser beams at 244 nm wavelength were employed 
to excite the n=2 state. The successful transitions were then detected  by 
photo-ionization  with  a  third  photon  from the same laser field. 
The released $\mu^+$ was then registered on a micro-channel plate detector.
\begin{figure}[ht] 
\centerline{\epsfxsize=4.1in\epsfbox{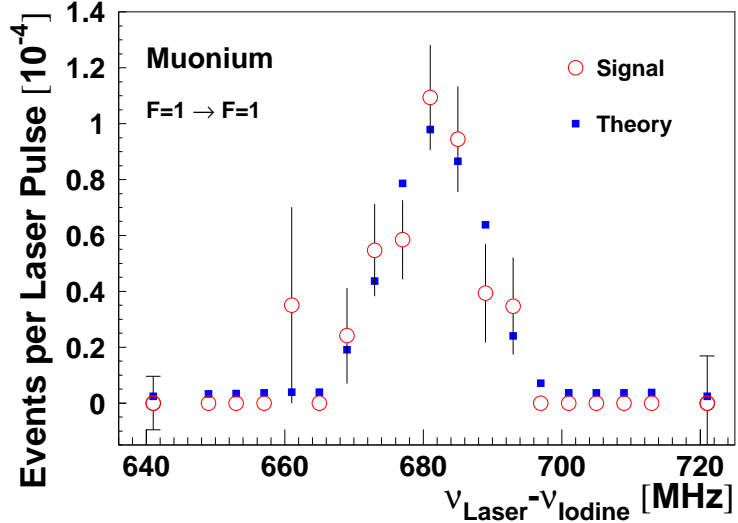}} 
\caption{\label{AAA} Muonium 1s-2s signal.  
         The frequency corresponds to the offset of the Ti:sapphire laser
         from the iodine reference line. 
         The open circles are the observed signal, the 
         solid squares represent the theoretical expectation 
         based on measured laser beam parameters
         and a line shape model.}
\end{figure}

\subsection{The recent RAL Experiment}

\begin{figure}[ht] 
\centerline{\epsfxsize=\textwidth\epsfbox{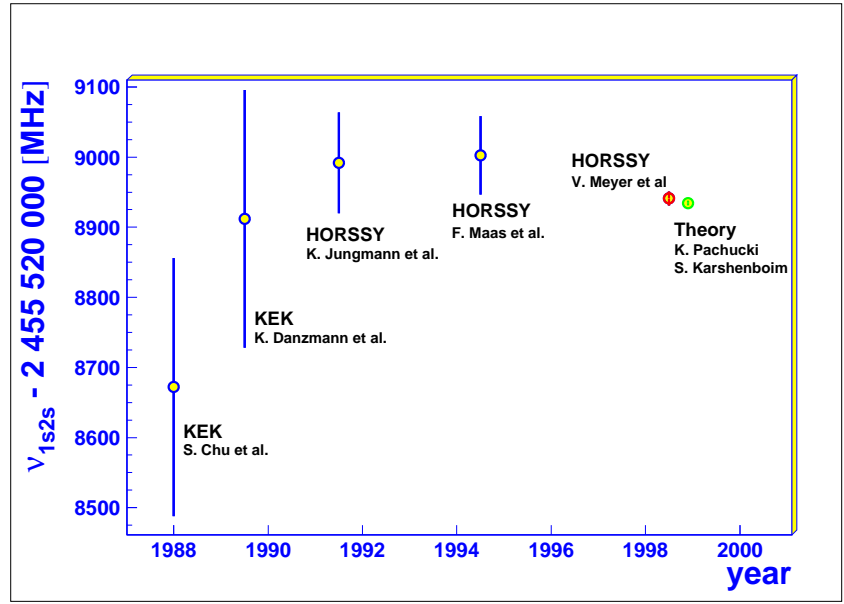}}   
\caption{\label{m1s2shistory} 
         Evolution of muonium 1s-2s measurements. The two results labeled KEK 
         accelerator facility refer
         to one single measurement by a Japanese - American collaboration 
         and its reanalysis. The newer measurements were
         made by the {\bf H}eidelberg -  {\bf O}xford - {\bf R}utherford - 
         {\bf S}ussex - {\bf S}iberia - {\bf Y}ale collaboration and have 
         reached 
         a level of accuracy comparable to theory, where the limitation arises
         primarily to the muon mass. }
\end{figure}

The accuracy of the early measurements was limited by the ac-Stark effect and 
rapid phase fluctuations (frequency chirps), which were inherent properties 
of the necessary pulsed high power laser systems (see Fig. \ref{m1s2shistory}).
The key feature for the 
latest high accuracy measurement 
at RAL was a shot by shot recording of the spatial laser intensity 
profile as well as the time dependences of 
the laser light intensity and phase. 
This together with a newly developed theory of resonant photo-ionization \cite{Yak_99} 
allowed a shot-by-shot 
prediction of the transition probability as a basis for the theoretical line 
shape (Fig. \ref{AAA}).
The latest RAL experiment \cite{Meyer_2000}
 yields $\Delta \nu_{1s2s}(exp)$= 2\,455\,528\,941.0(9.8) MHz in 
good agreement with a theoretical value $\Delta \nu_{1s2s}(theo)$=2\,455\,528\,935.4(1.4) MHz \cite{Pac_98}. 
The muon-electron mass ratio is found to be $m_{\mu^+}/m_{e^-}$ = 206.768\,38(17). 

Alternatively,
with  $m_{m^+}/m_{e^-}$
 as extracted from $\Delta \nu_{HFS}$, a comparison of experimental and 
 theoretical values for the 1s-2s transition 
can be interpreted in terms of a  $\mu^+$ -- $e^-$  charge ratio, 
 which results as  $q_{m^+}/q_{e^-} +1= -1.1(2.1) \cdot 10^{-9}$. This is the best verification 
 of charge equality in 
the first two generations of particles. The existence of one single universal 
quantized unit of charge  is solely an experimental fact and no underlying 
symmetry could yet be revealed. 
The interest in such a viewpoint arises because 
gauge invariance assures charge quantization only within one generation of particles.

\subsection{Slow Muons from Muonium Ionization}

A new development aims at providing low energy ($<$1~eV energy spread) 
muons for condensed matter research. At RIKEN-RAL facility a new
muon source is being set-up which bases on the laser photo-ionization
of muonium. The atoms are produced from hot metal foils which they
leave at thermal velocities. The ionization process involves 
one-photon excitation of the 1s-2p transition and subsequent ionization
with a second laser. At present the yield is a few $\mu^+$/s.
\cite{Nagamine_2003}

\subsection{Future Muonium Laser Spectroscopy}
 
Major progress in the laser spectroscopy of muonium can be expected from a continuous wave 
laser experiment, where frequency measurement accuracy does not present any problem
because light phase fluctuations are absent. For this an intense source of muons will 
be indispensable (see section \ref{Future}). which provides at least a factor of 1000 higher flux of (pulsed)
surface muons. As a promising 
starting point one would like to
have a pulsed beam
$5 \cdot 10^8$ $\mu^+$/s below 20 MeV/$c$ with typically 
3 \% momentum width and about 100 Hz repetition rate.

\section{Muonium and the Muon Magnetic Anomaly}

The muon magnetic anomaly $a_{\mu}$ (see contributions by
Francis Farley and Ernst Sichtermann to this volume \cite{Farley_2004})
is given, like in case of the electron,
mostly by virtual photon and electron-positron fields. However, the effects of heavier
particles can be
enhanced by the square of the muon - electron mass ratio $m_{\mu}/m_e \approx 4 \cdot 10^4$.

At the level of present experimental accuracy there are contributions to the muon magnetic anomaly
which are absent in the electron case.
For the muon the contributions of the strong interaction, which come in through vacuum polarization 
loops with hadronic content, can be determined using a dispersion
relation  and the input from experimental data on $e^+$-$e^-$ annihilation into hadrons and
hadronic $\tau$-decays. They  amount to 58 ppm. The weak interactions contributing through
W or Z boson exchange give a 1.3 ppm correction. 
At present standard  theory yields $a_{\mu}$ to about 0.7 ppm. 
Contributions from physics beyond the Standard Model
could arise from, e.g., supersymmetry, compositeness
of fundamental fermions and bosons, CPT violation and many other sources.
They could be at the ppm level.

The experimental values for the magnetic anomaly of  $\mu^+$ and $\mu^¯$ 
have been determined very recently
by a collaboration headed by Vernon Hughes at the Brookhaven national Laboratory (BNL).
It is a ''g-2'' experiment in which the 
difference of the spin precession and the cyclotron frequencies
is measured.
The experimental results for muons of both sign of electric charge 
are accurate to 0.7 ppm and agree well. Assuming CPT invariance they
yield a combined value of $a_{\mu}$ to 0.5 ppm \cite{Bennett_2004}. 
At this time it is unclear, whether there is
a small difference between theory and experiment at the level of 2 to 3 
standard deviations
due to unresolved issues in the theory of hadronic corrections. \\

\begin{figure}[ht] 
\centerline{\epsfxsize=\textwidth\epsfbox{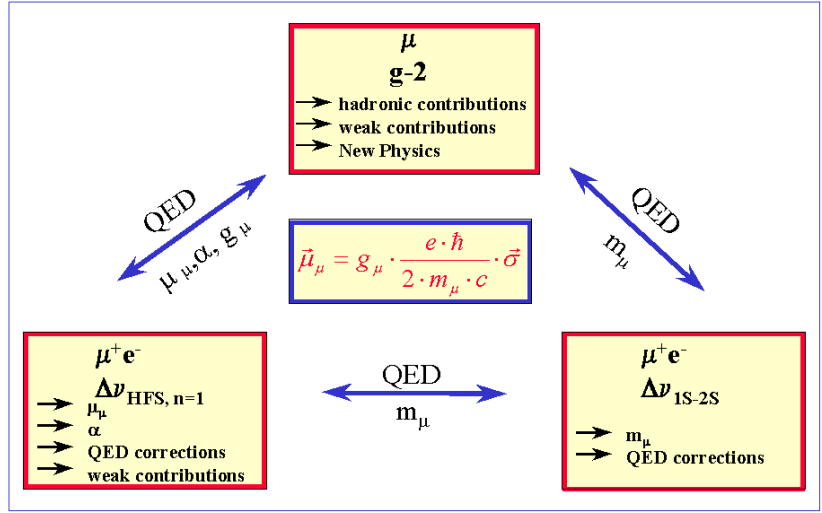}}   
\caption{The spectroscopic experiments on the hyperfine structure of muonium and the
         1s-2s energy interval are closely related to a precise measurement of the muon
         muon magnetic anomaly. The measurements put a stringent test on the internal
         consistency of the theory of
         electroweak interaction and on the set of fundamental constants involved.   }
\end{figure}

The microwave and laser spectroscopy of muonium are closely related
to the measurement of the muon magnetic anomaly. The fundamental constants
such as $m_{\mu}$, $\mu_{\mu}$, $\alpha$ and $q_{\mu^+}/q_{e^-}$ are 
indispensable 
input for the theory and the experiment on $a_{\mu}$. It should be noted that
prior to a future significant experimental improvement of  $a_{\mu}$, 
such as planned at the Japanese J-PARC accelerator complex
\cite{Carey_2003},  an improvement
in the knowledge of muon related fundamental constants would be required.
Muonium spectroscopy offers a clean way to obtain them.

\section{Muonium-Antimuonium Conversion}
\label{MMbar}

In addition to the indirect searches for signatures of new physics in the muon 
magnetic anomaly and in electromagnetic interactions within the muonium atom 
the bound state offers also the possibility to search more directly for 
predictions of speculative 
models. \\

\begin{figure}[ht] 
\centerline{\epsfxsize=4.in\epsfbox{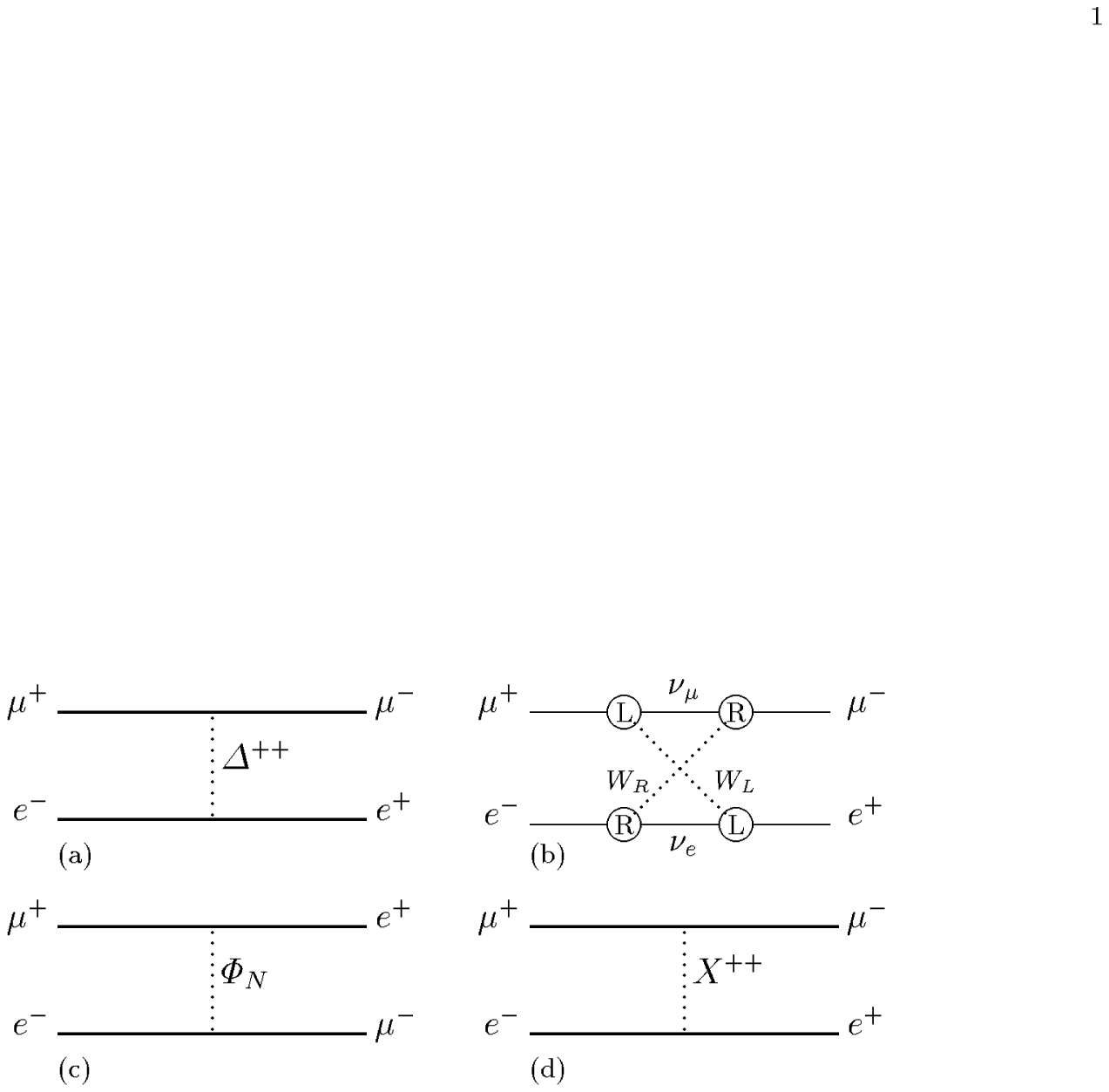}}   
\caption{\label{theo_mmb}
        Muonium-antimuonium conversion in
        theories beyond the standard model. The interaction
        could be mediated by
        (a) a doubly charged Higgs boson $\Delta^{++}$,
        (b) heavy Majorana neutrinos,
        (c) a neutral scalar $\Phi_N$, e.g.
        a supersymmetric $\tau$-sneutrino $\tilde{\nu}_{\tau}$, or
        (d) a bileptonic gauge boson $X^{++}$.  }
\end{figure}

The process of muonium to antimuonium-conversion 
(${\rm M}$-$\overline{{\rm M}}$)violates additive
lepton family number conservation. It would be an analogy in the lepton sector 
to  the well known $K^0$-$\overline{K^0}$ and 
$B^0$-$\overline{B^0}$ oscillations in the quark sector. Muonium-antimuonium 
conversion appears naturally in many theories beyond the Standard Model. 
The interaction could 
 be mediated, e.g., by a doubly charged Higgs boson $\Delta^{++}$, Majorana neutrinos, a 
 neutral scalar, a supersymmetric $\tau$-sneutrino  , or a doubly charged bileptonic
 gauge boson.

There have been a number of attempts to observe ${\rm M}$-$\overline{{\rm M}}$
conversion. The pioneering work was again 
performed by a group guided by Vernon Hughes
already in the 1960ies \cite{Amato_1967}. The early experiments 
\cite{Amato_1967,Huber_1990} relied
on the X-rays which would follow a $\mu^-$-transfer to a heavy element
upon contact of $\overline{{\rm M}}$ with matter as part of their signature.
A breakthrough
was the availability of thermal muonium in vacuum \cite{Woodle_1988}
which led to a significant increase in sensitivity \cite{Matthias_1991}. 

\subsection{ The latest ${\rm M}$-$\overline{{\rm M}}$ Experiment at PSI}

At PSI an experiment was designed to exploit a powerful new signature, which 
requires the coincident identification of both particles forming the anti-atom 
in its decay \cite{Willmann_1999}. The technique had been pioneered
by an international collaboration led by Vernon Hughes at LAMPF 
\cite{Matthias_1991}.
\begin{figure}[ht]
\centerline{\epsfxsize=3.2in \rotatebox{90}{\epsfbox{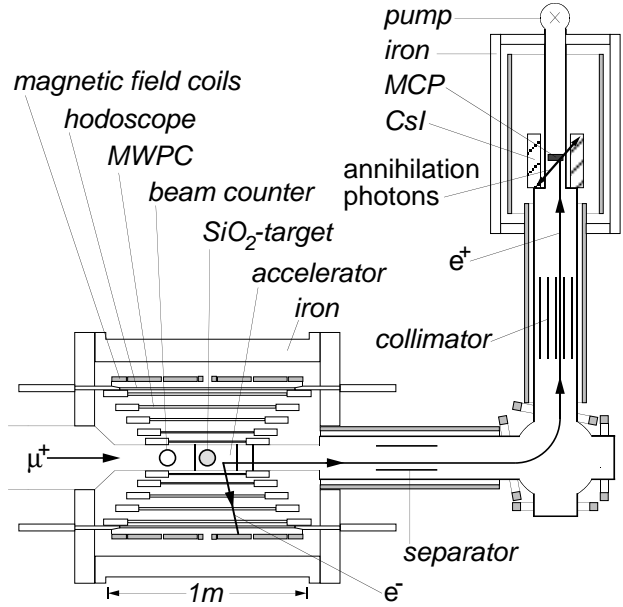}}}   
\caption{\label{exp_mmb}
        The Muonium-Antimuonium Conversion Spectrometer at PSI.   }
\end{figure}

Thermal muonium atoms in vacuum from a SiO$_2$ powder target, 
are observed for  decays. Energetic electrons from the decay of 
the $\mu^-$ in the   atom can be 
identified 
in a magnetic spectrometer (Fig. \ref{exp_mmb}). The positron in the 
atomic shell of $\overline{{\rm M}}$  is 
left behind after the decay with 13.5~eV average kinetic energy. It has been 
post-accelerated and guided in a magnetic transport 
system onto a position sensitive
 micro-channel 
plate detector (MCP). Annihilation radiation can be observed in a segmented 
pure 
CsI calorimeter around it. The decay vertex can be reconstructed. 
The measurements were performed during a period of 6 months in total over 4 
years during which $5.7\cdot 10^{10}$ muonium 
atoms were in the interaction region. 
One event 
fell within a 99\% confidence interval of all relevant distributions. 
\begin{figure}[ht] 
\centerline{\epsfxsize=2.1 in\epsfbox{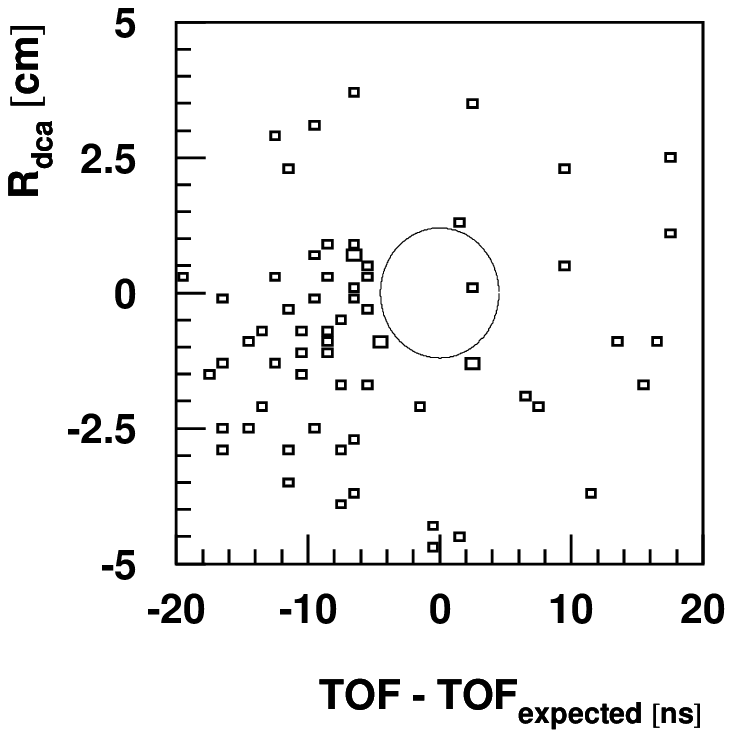}\epsfxsize=2.1in\epsfbox{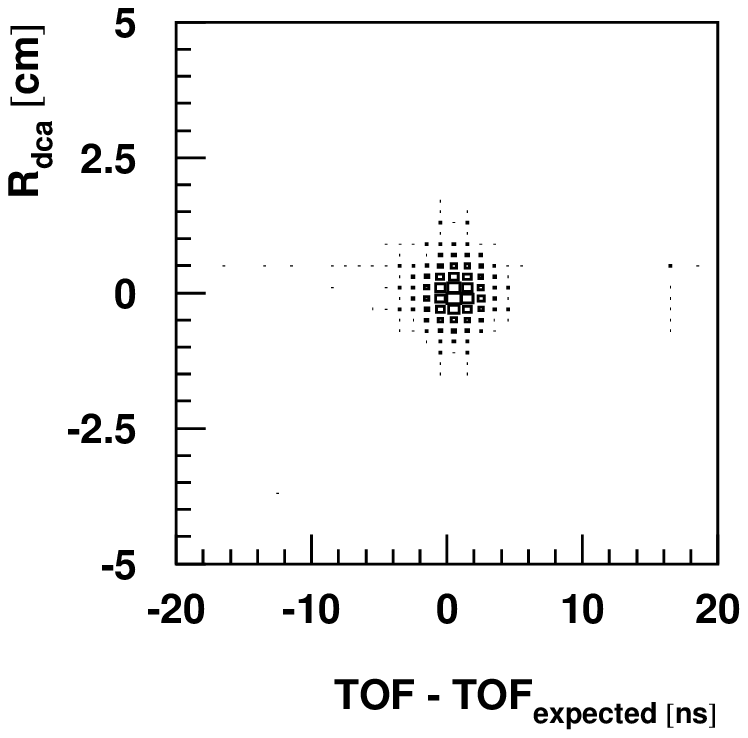}}   
\caption{\label{MMB_sig} Time of Flight (TOF) and vertex quality
        for the final 4 month of integrated search data for antimuonium
        (left) and for a single 20 minutes measurement of muonium (right).
        In the antimuonium case one event falls into the indicated area, 
        which corresponds to
        3 standard deviations for the parameters of an expected 
        antimuonium signature, which were 
        calibrated with muonium decays when all electro-magnetic 
        fields in the apparatus were reversed.
        }
\end{figure}
The expected background due to accidental coincidences is 1.7(2) events (Fig. \ref{MMB_sig}). Depending on
the interaction details one has to account for a suppression
of the conversion in the 0.1 T magnetic field
in the spectrometer. This amounts maximally
to a factor of about 3 for V$\pm$A type interactions. Thus, the upper limit
 on the conversion probability is $8.2 \cdot 10^{-11}$ (90\% C.L.). 
The coupling 
constant is bound to below $3.0 \cdot 10^{-3}~G_F$, where $G_F$ is the
weak interaction Fermi coupling constant. 

This new result, which exceeds limits from previous experiments by a 
factor of 2500 and one from an early stage of the experiment by 35, has 
some impact on speculative models. For example: A certain Z$_8$ model is ruled 
out. It had more than 4 generations 
of particles and where masses could be generated radiatively with heavy 
lepton seeding. A new lower limit of 2.6 TeV/$c^2 \times g_{3l}$ (95\% C.L.) on the 
masses of flavour diagonal bileptonic gauge bosons in GUT models is extracted, 
which lies well beyond the value
 derived from direct searches, measurements of the muon magnetic anomaly or 
high energy Bhabha scattering. Here, $g_{3l}$ is of order unity and depends on the 
details of the underlying symmetry. For 331 models the experimental result can 
be translated into  
 £ 850 GeV/$c^2 \times g_{3l}$ which excludes some of their minimal Higgs versions, where 
 an upper bound of 600 GeV/$c^2$ has been extracted from an analysis of electro-weak 
 parameters. The 331 models need now to refer to a less attractive and more complicated 
 extensions.
In the framework of R-parity violating supersymmetry the bound on the relevant 
coupling parameters could be lowered by a factor of 15 to  
$\lambda_{132} \cdot \lambda_{231} \leq 3 \cdot 10^{-4}$ for 
assumed super-partner masses of 100 GeV/$c^2$.

\subsection{Future Possibilities for ${\rm M}$-$\overline{{\rm M}}$ 
Experiments}
A future 
experiment to search for ${\rm M}$-$\overline{{\rm M}}$ conversion
could particularly take advantage of high intensity pulsed 
beams. In contrast to other lepton (family) number violating muon decays, 
the conversion 
through its nature as particle-antiparticle oscillation has a time evolution 
in 
which the probability for finding a system formed as muonium decaying as  
$\overline{{\rm M}}$ 
 increases quadratically
in time. This gives the signal an advantage, which grows in time over 
exponentially decaying background. E.g., with a twofold coincidence as part of
a signature: after a time $\Delta T = 2 \tau_{\mu}$  
beam related  accidental background has dropped by almost two orders of 
magnitude, whereas a  ${\rm M}$-$\overline{{\rm M}}$-signal would not have 
suffered significantly at all.
An almost ideal beam would 
have $1 \cdot 10^{10}$ $\mu^+$/s at below 23 MeV/$c$ with typically 
1-2 \% momentum width. The beam should be pulsed with up to
1$\mu$s wide pulses of up to several 10 kHz repetition frequency. 

Such efforts appear well motivated among other reasons 
through the connection
to numerous speculative models involving, e.g., 
lepton flavour violation in general or 
a possible  Majorana nature of the neutrinos \cite{Clark_2003}. 

\section{Future Possibilities for Fundamental Interaction Research with Muons}

\label{Future}
It appears that the availability of particles limits at present
progress in research with muonium. This includes better 
fundamental constants as well as the possibility to find
very rare processes
and to impose significantly improved limits in continuation of the search 
program of dedicated experiments.

Therefore significant  measures
to boost the muon fluxes at existing accelerator centers
and to create future facilities with orders of magnitude higher
muon currents
is indispensable prior to any significant progress.
This requirement matches well with the demand of several
communities within physics which request an intense
particle accelerator. Such interest exists, e.g., worldwide for 
a Neutrino Factory and a Muon Collider \cite{Alsharo_2003},
in Japan for the J-PARC facility and
in Europe for EURISOL facility \cite{Ridicus_2003}.
The perspectives for muonium research have beam worked out 
for such a scenario in some detail \cite{Alsharo_2003}.

Examples of tailored intense muon sources at existing facilities are     
the $\pi-\mu$ converter at PSI, the planned muon production of the
planned MECO experiment at BNL or the projected phase rotated intense 
source of muons (PRISM) \cite{Kuno_2003} in connection with the upcoming 
J-PARC facility.

The design goal for such a new machine should be
a minimum of 1 MW proton beam on a production target.
This could become a reality at various places
including BNL, CERN, GSI and J-PARC.
\begin{figure}[ht] 
\centerline{\epsfxsize=4.in\epsfbox{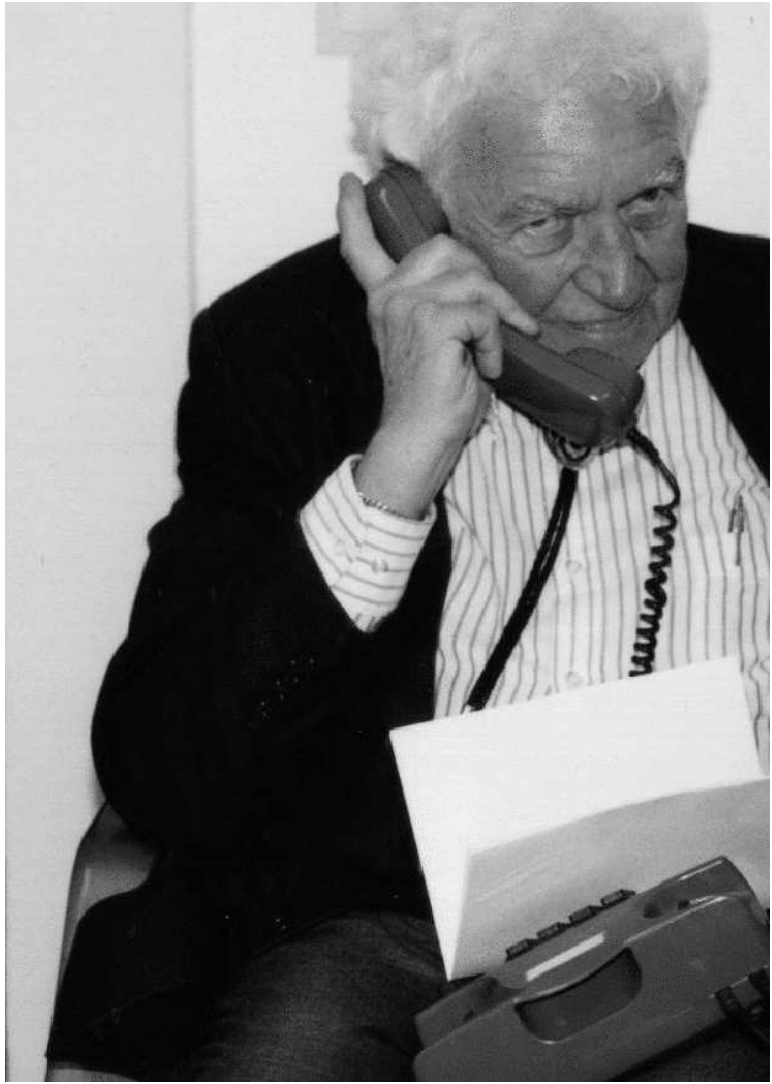}}   
\caption{Vernon W. Hughes keeping contact with his colleagues 
during the  Lepton Moments I conference in Heidelberg in 1999.
Effective communication is a key to operate simultaneously
several challenging large experiments at different research 
centers around the world.}
\end{figure}

\section{Conclusions}

More than 70 years after the muon was  discovered \cite{Kunze_1933},
its properties still remain puzzling.
The mysteries include not less than the reason for the muon's existence,
the size of its mass, its charge and the fact of
lepton number (and for charged species) lepton family
number conservation. All precision experiments
to date have confirmed that the muon is a heavy copy
of the ''point-like'' electron. Standard theory appears to
be an adequate description of all measurements, although it
cannot answer some of the deeper questions.

Searches for a violation of the Standard Model were not blessed 
with a successful observation yet. However,
both the theoretical  and experimental work in this connection have led
to a much deeper understanding of 
particle interactions. One special value of the precision
experiments are their continuous contributions towards
guiding theoretical developments by excluding various speculative models.
The research of Vernon Hughes has most significantly added 
to shaping our knowledge about nature in this way. Research on muonium
has contributed quite some essential facets to the
picture of fundamental particles and fundamental interactions in
physics.

\section{Final Remarks} \footnote{The author wishes to thank C.J.G. Onderwater
for carefully reading the manuscript and his help during formatting.}

The author owes his heartily gratitude to Vernon Willard Hughes
for a long fruitful collaboration in the framework of a number of
international collaborations. The measurements we could carry out together
all involved muons.  In particular, many of them involved centrally
the fundamental atom we could find for our research:

\noindent{\it Thank You Vernon for introducing muonium to the scientific community.}


\end{document}